# Phase stochastic resonance in coherent QKD signal transmission with a crosstalk noise in multicore fiber

E Ponizovskaya-Devine[1]

(Dated: August 17, 2022)

The study considers the new effect of phase stochastic resonance in a linear system with multiple external signals that can influence quantum communication. The phase stochastic resonance was shown in a linear co-propagation of the quantum and multiple classical signals in the multicore fiber of the quantum channel of Quantum Key Distribution (QKD). The dependence of SNR vs phase noise of classical signal could have a maximum at a certain range of phase noise depending on the wavelength of the signals. The results determine the set of transmission parameters that would produce less interference with the quantum signal in QKD systems.

## I. INTRODUCTION

Stochastic resonance is a phenomenon that still produces wide interest both as a scientific phenomenon and for its applications in secure communications. Recently, many studies were published on stochastic resonance in bistable systems [1, 2]. The study is considered an interesting and new phenomenon that shows the stochastic resonance in quantum communications networks, which is very important for network design. The quantum network proposes the solution for very secure communication and recently are a hot topic. One example is the Quantum Key Distribution (QKD) system [3,4,5]. The secure communication method implements a cryptographic protocol that enables two parties to produce a shared random secret key known only to them and is used to encrypt and decrypt messages. An important and unique property of QKD is the ability of the two communicating users, Alice and Bob, to detect the presence of any third party trying to gain knowledge of the key. The bits are encoded with a quantum state carried by single photons (qubits). There are different ways to encode qubit values on single photons. One of the ways to encode qubits used in the QKD system is time-bin qubits that create a pair of coherent pulses propagating in the same spatial mode and separated by a given time. The noise in quantum communication and quantum processors is an important problem that can influence the performance [6] and could be crucial for the design. In the study, the question is how much electromagnetic interference could be handled if a quantum signal is propagating in a multicore fiber (MFC) with standard crosstalk with neighboring cores and find the first time reports the SR phenomena. The MFC is available with crosstalk at about -40-50dB [7]. Another important question is to optimize the communication line to increase the signalto-noise ratio (SNR).

According to COW protocol [5] there is the first or early pulse and the second one is called the late pulse. The two quantum states compose the computational basis of the qubit space. The implementation of a time-bin qubit emitter is based on an unbalanced Mach-Zehnder interferometer where the input beam splitter ratio can be varied and the output beam recombiner is a fast switch. The BB84 protocol is implemented with time-bin qubits in Clavis3 [5] as four qubit states located on the equatorial plan of the qubit sphere. The two Mach-Zehnder interferometers (MZI) are made with two 50/50 couplers and one phase modulator. The implementation requires tight control of the interferometer's stability. The study of this paper aimed to find the requirements of the crosstalk in the quantum channel. Theoretically analyzing environmental influence the following conditions of COW protocol is considered (Fig.1a): an emitter at Alice emitting qubits states from the computational basis or decoy sequences; the time between all consecutive pulses is identical and the phase relation between those consecutive pulses is kept constant; the ratio of the number of qubits from the computational basis and the number of decoy sequences is in favor of the computational basis; the receiver station analyse the computational basis and check the phase relation between two consecutive optical pulses; the ratio of use of the analyzer for the computational basis compare to the use of the analyzer for the phase relation check is in favor of the computational basis; A QBER value is measured by counting the probability of having an error in the exchange of qubits of the computational basis; the phase relation check is quantified by measuring the visibility of interferences occurring in the second basis analyzer; Based on both values, QBER and visibility, it is possible the estimate if it is possible to extract secret keys form the qubits exchanged between the emitter and the receiver stations. The system records the quantum bit error (QBER), introduced by the external signal from crosstalk, and Visibility(Fig,1a). The external signal repeats the path of the quantum signal through Mach-Zehnder Interferometer to the detectors. The signal is characterized by amplitude, and wavelength, which can be different than the wavelength of the qubit and the phase noise. The system includes MZI, which resonates at a set

[1] University of California, Electrical and Computer Engineering Davis Davis, USA ; eponizovskayadevine@ucdavis.edu

of wavelengths depending on the delay and has a set of resonance frequencies.

Stochastic resonance (SR) is a nonlinear effect, with the unique property that the signal/noise ratio (SNR) increases with increasing noise and reaches a maximum at a quite high noise level. It usually occurs in bi-stable systems and is associated with a noise-induced transition from one state of the system into another [8-11]. Recently, SR was observed for the phase noise [2]. The effect was called the phase stochastic resonance (PSR). Although SR is often associated with nonlinear systems it was observed in a linear system with external signals and noise [12]. In this case, the phase noise can induce the transformation between the MZI resonance frequencies increasing or decreasing SNR for a certain range of the phase noise. Choosing the noise range we consider that the coherent laser usually has a linewidth of about only 1-100Hz. However, a short femtosecond pulse that is normally used for communication could have a linewidth in the range of THz. In the current study, the experiment was done to show that the long coherent pulses could disrupt communication experimentally. The theoretical model was developed and validated for narrow noise band phase noise and the results for a wider linewidth noise reviled the SR. The external signal that consists of pulses as short as femtoseconds could have less effect on the quantum channel and the SNR for the quantum channel has a maximum for a certain range of phase noise depending on the frequency. In the environment of several pulses, the SR could be observed.

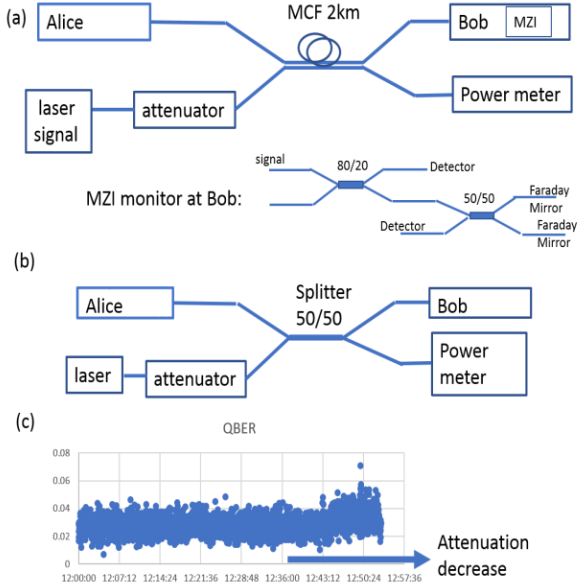

FIG. 1. The schematics of QKD with MCF: a) the crosstalk through 2km fiber influence on QKD experiment, insert – schematics of the QKD monitor (b) experimental setup for the influence of the directly injected signal with noise into a quantum channel, (c) QBER recorded with crosstalk

## II. THEORETICAL MODEL

First, the simple model of the system was developed and tested experimentally. The basic assumptions for the model are: QKD signal going through MZI produces a signal at detectors with probability $D_1$ and $D_2$ taking into account that the MZI delay is chosen to produce 0 at $D_1$ in the ideal case; crosstalk from external laser signal also goes through the MZI, producing the signal at $D_{m1}$ and $D_{m2}$ that is registered in the same time frame as QKD; the laser noise is assumed to have certain statistics and laser signal experience a random phase shift at MZI due to it. At Bob site there is monitor with MachZehnder interferometer (MZI) with photodetectors $D_{m1}$ and $D_{m2}$. The laser signal measure during time $\Delta T$ produces a click at the detectors with probabilities:

$$D_{m2} = \cos^2(\theta/2)$$
$$D_{m1} = \sin^2(\theta/2) \quad (1)$$

The visibility was determined as

$$V = \frac{p(D_{m2}) - p(D_{m1})}{p(D_{m2}) + p(D_{m1})} \quad (2)$$

Where $p$ denotes the conditional probability to observe a click in $D_{mi}$. We assume that $D_1$ and $D_2$ are the probability of the clicks if we have only a QKD signal. We assume that we have injected an external signal with some frequency difference compared to Alice source $\Delta\omega$. We know that the delay time of MZI is $\Delta t$ (50ps according to the manual), which gives us a phase shift of $\theta_\omega = \omega \Delta T/2$. The phase noise of the signal is characterized as a frequency f distribution with mean value $\langle f_n \rangle = 0$ and some mean square $\sigma^2 = \langle f_n^2 \rangle$. The externally injected laser signal propagates to Bob through the same MZI creating an additional current at $D_{m1}$ and $D_{m2}$. As we have noise we have to average the results by noise sampling. That means the additional terms at the detectors are:

$$p(D_{m2}) = D_2 + \alpha P_l \langle \cos^2(\theta/2 + 2\pi f_n \Delta T) \rangle \quad p(D_{m1}) = D_1 + \alpha P_l \langle \sin^2(\theta/2 + 2\pi f_n \Delta T) \rangle \quad (3)$$

Brackets here mean averaging, $P_l$ is the external laser power and $\alpha$ is a normalization coefficient related to the amplitude of the quantum signal. In the case of the interference from the multiple cores with signals with different amplitude, frequency, and noise bandwidth it will be a sum over all cores

The visibility with several crosstalk signals from multiple cores $i$ can be defined as

$$V = \frac{D_2 - D_1 + \sum_i \alpha P_{li} \langle \cos(\theta_i + 4\pi f_{ni} \Delta T) \rangle}{D_2 + D_1 + \alpha P_{li}} \quad (4)$$

Assuming that the noise is small we can expend cos and sin in the Taylor series and keep only second-order terms. Eq(4) shows that the system can exhibit nontrivial dependence on phase noise that will be studied here.
First, we see the results for the low phase noise and compared them with the experiment. The probability of the click in the case of QKD and laser one signal is

$$p(D_{m2}) = D_2 + \alpha P_l/2 + \alpha P_l V_\omega * \left(1 - (\pi\sigma\Delta T)^2\right)/2$$
$$p(D_{m1}) = D_1 + \alpha P_l/2 + \alpha P_l V_\omega * \left(1 - (\pi\sigma\Delta T)^2\right)/2 \quad (5)$$

In this case, visibility is

$$V = \frac{D_2 - D_1 + \alpha P_l V_\omega - \alpha P_l V_\omega (\pi\sigma\Delta T)^2}{D_2 + D_1 + \alpha P_{li}} \quad (6)$$

$V_\omega$ is the term due to frequency mismatch.

$$V_\omega = \cos^2\theta_\omega - \sin^2\theta_\omega \quad (7)$$

The result shows that different wavelengths of the external laser effects visibility differently. $-1 \leq V_\omega \leq 1$ The minimal $\Delta\lambda$ that shifts the phase to $2\pi$ is $\Delta\lambda = 2\lambda^2/(c\Delta T - 2\lambda)$ $2\lambda^2/(c\Delta T)$, $c$ – velocity of light, $\lambda$ – wavelength, if $\Delta T = 50ps$, $\Delta\lambda$ is 0.3 nm For one pulse it must be impossible to distinguish between two lasers with different but very small noises less than MHz. If the phase noise is negligibly small the visibility can be estimated as follows: If the phase noise is negligibly small the visibility can be estimated as follows:

$$V = \frac{D_2 - D_1 + \alpha P_l V_\omega}{D_2 + D_1 + \alpha P_{li}} \quad (8)$$

This expression was used to fit the data. The model parameters were taken from the experiment with the continuous laser signal with low phase noise injected into a quantum channel at Spectrum Lab, MSU.

### III. EXPERIMENT AND MODEL VALIDATION

A set of experiments were conducted to estimate the influence of the classical signal propagating in the multicore fiber (MCF) next to the quantum channel (QC). The Quantum Key Distribution (QKD) system is an example of communication where the performance of the QC can be interrupted by the crosstalk if quantum and classical signals co-propagate in the cores of the same MCF. The possible crosstalk in the MCF was measured and found to be within typical values between -40-50dB. The measurements for crosstalk in 2km fiber also confirmed that the crosstalk is less than -40dB. The influence of crosstalk was modeled experimentally by producing the direct injection of an external laser in QC through the attenuator that reduces the signal to values similar to the crosstalk values (Fig.1b).

The QKD system Clavis-3 [5] was used with the communication between Alice and Bob provided through a 2km multicore fiber. The setup is shown in Fig.1 (a). It consists of the QC between Alice and Bob and includes the 2 km 4-core MCF fiber. One of the cores is used for QC and another transmits a signal from a laser that goes through the attenuator and is measured by the power meter on the other end of the MCF fiber. NKT, Pure Photonics (PP) laser in a whisper and in normal modes, and RIO laser were used in the experiment, they are all low noise, however, PP laser had a slightly wider noise bandwidth (about 10-15KHz) while RIO and NKT noise bandwidth was only 1-2kHz. As the crosstalk was small we set the attenuator for 0dB. A power meter was used on the free end of the 50/50 splitter to control the laser power. QBER and Visibility were measured on the Clavis-3 system. Usually, QBER is about 2%. It sharply increased after we start transmitting the laser signal and then stabilized at 4% (Fig.1c). PP laser had a variable wavelength in a range from 1530 to 1560 nm. The QKD system transmits its signal at 1550nm wavelength and has a bandpass filter at about 1550nm. We changed the wavelengths and have seen the response in QBER that is shown in Fig2a. Each frequency stays for 5 min then the laser is disabled, the frequency switched to the next, and enables again. The Key generation never was interrupted although we noticed an increase in QBER. We simplified the experiment by producing the direct injection of an external laser in QC through the attenuator to reduce the signal to values similar to the crosstalk values.

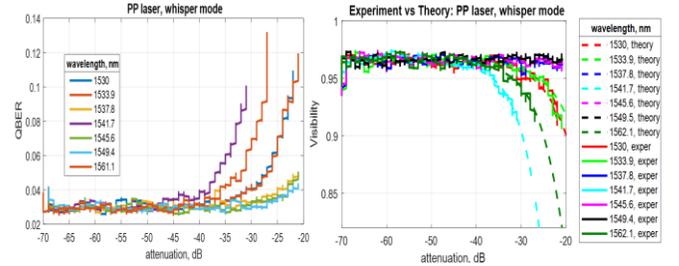

FIG. 2. The injected external signal with QKD experiment for PP laser in whisper modes: (a) QBER vs attenuation for whisper mode, (b) visibility vs attenuation for whisper mode compared with the model Eq.8.

The direct injection data was consistent with the crosstalk data that showed the effect starting at about the power -40-50dB and the key distribution can be disrupted at the higher power. However, the threshold was 11-15dB lower for RIO and NKT lasers as well as for certain wavelengths at PP lases. It could be a result of either phase noise or different wavelengths. However, the difference in noise was too small to cause the results. The data collection for QBER and Visibility vs attenuation for PP laser whisper mode is shown in Fig.3. The dashed lines in

Fig.3b show the theoretical visibility from Eq.8. $D_1$ and $D_2$ are taken from the Visibility of the QKD system without an external signal. The parameter $\alpha$ was fitted and was found to be 15dB. The same fitting parameters were valid for other measurements. The experimental data fit the model. Wavelength shift as low as $\Delta\lambda = 0.3nm$ changes $V_\omega$ from plus to minus it can switch the threshold from -40dB (light blue curve) to less than -10dB (black, magenta, dark blue curves). The effect of the narrow filter at QKD Bob entrance showed as the increase of parameter $\alpha$ for the wavelengths near 1550nm, although the accuracy of the experiment allowed us to see it only for positive $V_\omega$. The phase noise was too small to affect the results significantly.

## IV. SR SIMULATIONS AND DISCUSSION

SNR was defined as the QKD signal power to the total external signal power, that is the parameter $(\alpha P_l)^{-1}$ in the Eq (4), at that has Visibility at a threshold 80%. Below the threshold value, the key can't be generated.

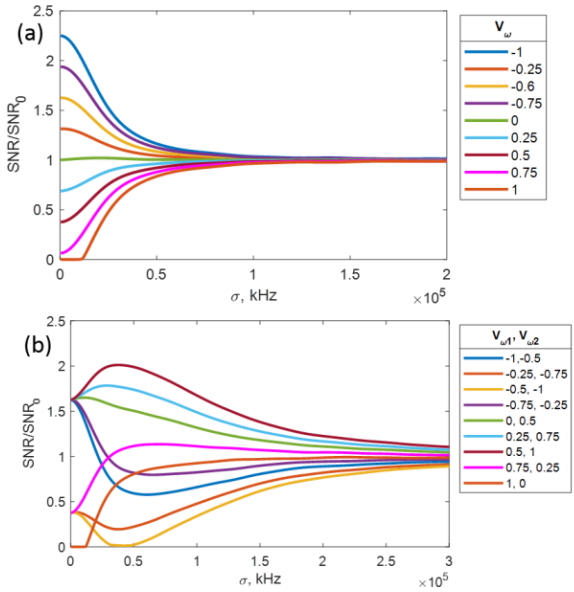

FIG. 3. SNR vs phase noise bandwidth: a) with one external signal, b) with two external signals showing the phase stochastic resonance.

The classical signal can have a noise bandwidth from a few kHz for a high coherent laser to THz for short femtosecond pulsed. We estimated theoretically the dependence of the noise for crosstalk from one external signal depending on its phase noise using Eq.(4), simulating the noise with the bandwidth and averaging the results over time. The results show that for the high noise the SNR is a constant value. SNR normalized to the SNR for the infinite bandwidth is shown in Fig. 3a vs the phase noise bandwidth for several $V_\omega(\lambda)$ values. As one can see, the SNR can go to zero for several external signal wavelengths, while it has a maximum at the others, and it goes to a constant value with the increase of the phase noise of the external signal. At a high level of phase noise, the MZI is not that sensitive to the external signal. However, the right frequency of the external signal can increase the SNR and prevent the QKD system of been affected by the crosstalk. An interesting result of the simulations for the one external signal is that the phase noise of the external signal can decrease the SNR for several resonance wavelengths, or bring it back from 0 to a relatively high number (see Fig.3a for positive $V_\omega$). A more interesting dependence of the noise happens if we have several signals crosstalk. Assume two external signals have different noise bandwidths and different wavelengths so that the ratio between the noise bandwidth of two signal crosstalk is a constant parameter. In this situation, instead of the case of noise-induced fluctuation between two stable states for SR [10], we have two external signals with different phase noise characteristics that switch the system between resonance states and affect the SNR similar to the SR in the linear resonance system with external noise [12]. The Fig.3b shows SNR vs noise bandwidth ($\sigma$) of the first signal while the second bandwidth is $0.2\sigma$ for the different sets of $V_{\omega 1}$ and $V_{\omega 2}$. In some cases, we see a phase stochastic resonance when the SNR increases at the low noise, reaches the maximum, and then decreases to the high noise level. There is a maximum of SNR at a certain phase noise in GHz that indicates a phase SR. The maximum shifts with the wavelength of the external signal and can disappear depending on external signal parameters. As we can see the correct configuration can increase SNR, while the wrong configuration can completely reduce it to 0 disrupting QKD from functioning.

## V. CONCLUSIONS

The study showed the effect of phase stochastic resonance for a quantum channel of a QKD system with two external signals with phase noise. The stochastic resonance occurs due to noise-induced transformation between resonance wavelengths of the Mach-Zehnder interferometer. As a result of it, the SNR at a given wavelength of the external signals has a maximum for a certain range of noise. The results can be interesting for the design of quantum communication systems and put restrictions on the signal parameters for the optimal configuration.

## VI. ACKNOWLEDGMENT

I acknowledge Montana State University, Spectrum Lab which provided the QKD system for the experiments. The work was partially funded by AFRL in the experimental part.